\documentclass{aastex631}

\begin{document}

\title{Auroras on planets around pulsars}

\author[0000-0002-9475-5193]{Ruchi Mishra}
\affiliation{Nicolaus Copernicus Astronomical Center of the Polish Academy of Sciences,\\ Bartycka 18, 00-716
Warsaw, Poland}

\author[0000-0002-3434-3621]{Miljenko \v{C}emelji\'{c}}
\affiliation{Nicolaus Copernicus Astronomical Center of the Polish Academy of Sciences,\\ Bartycka 18, 00-716
Warsaw, Poland}
\affiliation{Research Centre for Computational Physics and Data Processing, Institute of Physics, Silesian University in Opava, Bezru\v{c}ovo n\'am.~13, CZ-746\,01 Opava,
Czech Republic}
\affiliation{Academia Sinica, Institute of Astronomy and Astrophysics, P.O. Box 23-141,
Taipei 106, Taiwan}

\author[0000-0002-6114-0539]{Jacobo Varela}
\affiliation{Universidad Carlos III de Madrid, Leganes, 28911, Spain}

\author[0000-0003-3095-6065]{Maurizio Falanga}
\affiliation{International Space Science Institute, Hallerstrasse 6, 3012 Bern, Switzerland}
\affiliation{Physikalisches Institut, University of Bern, Sidlerstrasse 5, 3012 Bern, Switzerland}

\begin{abstract}
The first extrasolar planets were discovered serendipitously, by finding the slight variation in otherwise highly regular timing of the pulses, caused by the planets orbiting a millisecond pulsar. In analogy with the Solar system planets, we predict the existence of aurora on planets around millisecond pulsars. We perform the first magnetohydrodynamic (MHD) simulations of magnetospheric pulsar-planet interaction and estimate the radio emission from such systems. We find that the radio emission from aurora on pulsar planets could be observable with the current instruments. We provide parameters for such a detection, which would be the first radio detection of an extrasolar planet. In addition to probing the atmosphere of planets in such extreme conditions, of great interest is also the prospect of the first direct probe into the pulsar wind.  
\end{abstract}

\keywords{ Pulsar-Planet -- MHD -- Aurora}

\section{Introduction} \label{sec:intro}
All active stars have magnetized wind \citep[and references therein]{Mestel68, Bouvier14}. When this magnetized flow is obstructed by the presence of an obstacle such as a planet or its moons, it leads to a partial dissipation of the energy flow \citep{Varela16}. The radiation emitted from this dissipation can be used to study star-planet magnetospheric interaction, which is governed by the stellar wind (SW), interplanetary magnetic field (IMF), and the intrinsic magnetic field of the planet \citep{See14}. In particular IMF, the component of the stellar magnetic field that is carried into interplanetary space by the stellar wind, modifies the intrinsic magnetosphere of a planet \citep{Blanc05}. Most of the power in such interaction is emitted as electromagnetic radiation in the visible range, while a fraction of it is emitted as cyclotron radio emission. Planets with intrinsic magnetic fields emit at low frequency radio wavelengths, through emission produced by the electron–cyclotron maser instability (CMI) \cite{Zarka98}. It is due to the energetic electrons traveling along the magnetic field lines which are formed in the reconnection region between the planetary magnetic field and IMF. This phenomenon is usually observed in aurora regions \citep{1979W, Strugarek19}.

There are a number of observations and case studies of such radio emissions in the Solar system, where most of the planets as well as some of the moons of the planets, exhibit aurora \citep{aurora2015}. When we consider low-frequency radio range, for all magnetized planets in the solar system (Mercury, Earth, Jupiter, Saturn, Uranus, and Neptune), this radiation is only up to 2 orders of magnitude less intense than the radiation produced by the Sun \citep{Zarka07}.
Aurora-like emissions are visible in the Solar system also near the planets without intrinsic magnetic fields, Venus and Mars \citep{Blanc05}. During the increased solar activity, Venus often shows a green glow in the magnetotail-the exact mechanism for this emission is still unknown \citep{GrayVenus14}. Mars shows two kinds of proton aurora, continuous and patchy, resulting from the interaction of incoming particles with its day-side atmosphere \citep{Deigh18, Milby20}.  

 The first extrasolar planets were serendipitously discovered around a millisecond pulsar B1257+12 \citep{firstexo}\footnote{The possibility of planets around a bright pulsar PSR B0329+54 was first suggested in \cite{Demianski79} but was not confirmed by observations at the time, and is still a tentative case \citep{Starovoit17}.}. It has three planets: one with $0.02 M_\oplus$ and other two super-Earth planets with $4 M_\oplus$ \citep{1994Wolsz}.
 
 A large majority of pulsars are isolated: only a fraction, less than 0.5\%, have planetary or planet-like companions \citep[see e.g.,][]{Nitu22}, and their characteristics vary from asteroid-dimensions or small rocky planets to gaseous supergiants. 
Radio emission from planets and small objects in the pulsar wind was first studied in \cite{Mottez11a,Mottez11b,Mottez20cor}, extending the theory of Alfv\'{e}n wings to the relativistic regime, and examining the influence of the related current system on the evolution of their orbit. In \cite{Mottez14,Mottez20}, a relativistic aberration narrowing the beam of emission from such wings nearby the smaller body orbiting a pulsar was evoked as a possible explanation for Fast Radio Bursts (FRBs).

Here we perform simulations of pulsar-planet magnetospheric interaction, to quantify the emitted electromagnetic radiation from pulsar planets. In Sec.~\ref{numsims} is presented our setup in PLUTO code for exploration of the radio emission generated from the planets orbiting around pulsars. The results from our simulations are presented in Sec.~\ref{plplan}, where in Sec.~\ref{radioemis} we compute the expected radio emission.

\section{Pulsar-planet magnetospheric interaction}\label{numsims}
\begin{figure}
\plotone{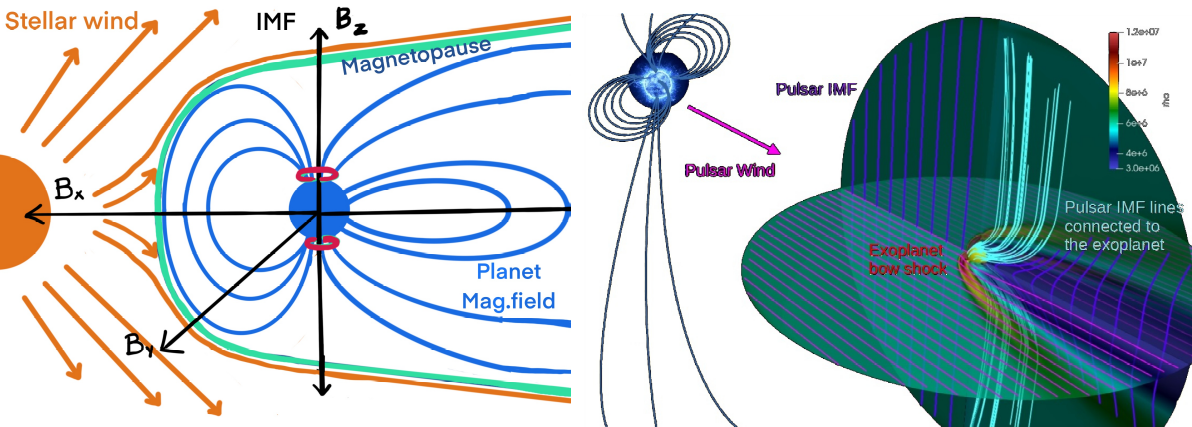}
\caption{{\it Left panel}: the schematic representation of a star-planet system with the magnetospheric interaction. The orange lines indicate stellar wind, blue lines indicate the planetary magnetosphere, and black vectors indicate the different components of the IMF. The aurora emission region is indicated near the  poles of the planet with red lines. Due to the magnetospheric interaction, planetary magnetic field lines are squeezed at the day side, and elongated at the night side. The boundary between magnetosphere and surrounding plasma is a magnetopause. {\it Right panel}: three-dimensional view of a pulsar-planet system in our simulation with conducting planet surface. The density distribution is shown in a color scale, induced magnetic field lines near the planet dayside, arising because of the piling-up of the IMF lines are shown with cyan lines, and IMF with blue lines. The direction of the pulsar wind is represented by the violet arrow. The simulation domain is inside the spherical region indicated by the intersected coloured circles. The pulsar itself is not a part of the simulation, its location is assumed beyond the outer boundary. }
\label{schexo}
\end{figure}
\begin{figure}[ht!]
\plotone{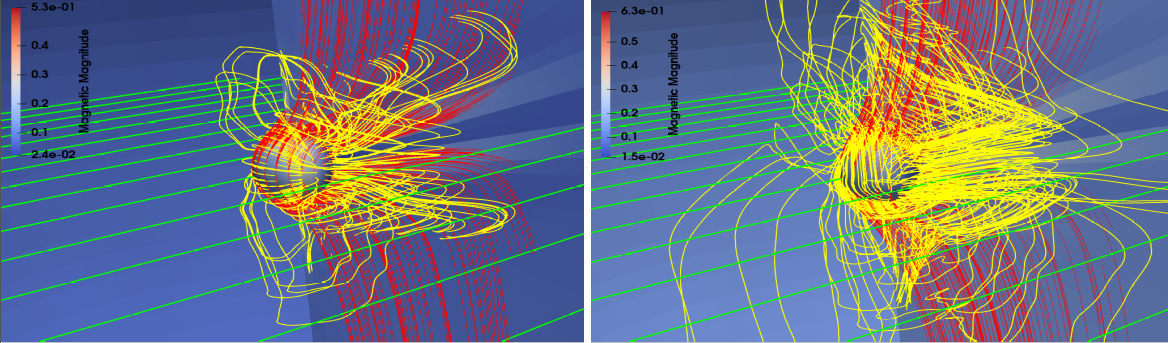}
      \caption{Results in our simulations with pulsar planets without intrinsic planetary magnetic field. With yellow lines are shown the electric currents, red lines are the magnetic field lines connected to the planet's atmosphere and green lines show the velocity streamlines of pulsar wind. Magnetic field magnitude is shown in color grading. {\it Left panel}: for the conducting planet's surface, electric current loops remain close to the planet's surface. {\it Right panel}: for the ferromagnetic planet surface, the currents show an extended dipolar electric field structure.}
         \label{simforsce3}
   \end{figure} 
Here we discuss numerical simulations of pulsar-planet magnetospheric interaction in three dimensions with the PLUTO code \citep{Mignone07}. The numerical model and setup we use are almost identical to the setup in a series of works by \citet{Varela16a, Varela16, Varela22}, presented in detail in \cite{Varela18}. The main difference is the use of special relativistic module in the code. Also, at the inner boundary, we assume a non-magnetic planet and change the conditions accordingly, as described below.

The main equations solved by the special relativity ideal MHD module of PLUTO, which we use here, are the conservation laws:
\begin{equation}
  \frac{\partial}{\partial t}\left(\begin{array}{c}
  D  \\  \noalign{\medskip}
   \mathbf{m}  \\  \noalign{\medskip}
 E_{\mathrm t} \\  \noalign{\medskip}
  \mathbf{B} \end{array}\right) 
  +\nabla\cdot\left(\begin{array}{c}
  D\mathbf{v} \\  \noalign{\medskip}
   (\rho h+{b^2_{\mathrm m}})\gamma^2 \mathbf{v}\mathbf{v}-\mathbf{b}\mathbf{b}+\mathbf{{\mathbf I}}p_{\textrm t} \\  \noalign{\medskip} \mathbf{m}\\ 
   \noalign{\medskip}\mathbf{v}\mathbf{B}-\mathbf{B}\mathbf{v} \end{array}\right)^T=\left(\begin{array}{c}
  0  \\  \noalign{\medskip}
  \mathbf{f}_{\mathrm g} \\  \noalign{\medskip}
     \mathbf{v}\cdot\mathbf{f}_{\mathrm g} \\  \noalign{\medskip}  \mathbf{0} \end{array}\right)\;,
\end{equation}
with the Lorentz factor $\gamma=(1-|\mathbf{v}|^2)^{-1/2}$, where $\mathbf{v}$ is the matter velocity (normalized to the speed of light, c) and D=$\rho\gamma$ represents the laboratory density. The specific enthalpy h is dependent on thermal pressure p and matter density $\rho$: h=h$(p,\rho)$. The momentum density is $\mathbf{m}=(\rho h\gamma^2+|\mathbf{B}|^2)\mathbf{v}-(\mathbf{v}\cdot\mathbf{B})\mathbf{B}$,$\mathbf{B}$ is the magnetic field, E$_{\mathrm t}$ is the total energy with the contribution from the rest mass so that $E=\rho h\gamma^2-p+|\mathbf{B}|^2/2+(|\mathbf{v}|^2|\mathbf{B}|^2-(\mathbf{v}\cdot\mathbf{B})^2)/2$ and the acceleration term $\mathbf{f}_{\mathrm g}=\rho\gamma^2[\gamma^2\mathbf{v}(\mathbf{v}\cdot\mathbf{a})+\mathbf{a}]$, where $\mathbf{a}$ is the acceleration. The total pressure is a sum of thermal and magnetic pressures $p_{\mathrm{t}} = p + b_{\mathrm{m}}^2/2$, and $b^2_{\mathrm{m}}=|\mathbf{B}|^2/\gamma^2+ (\mathbf{v}\cdot\mathbf{B})^2$, $\mathbf{b}=\mathbf{B}/\gamma+\gamma (\mathbf{v}\cdot\mathbf{B})\mathbf{v}$ \citep{Mignone07}.

The above set of conserved equations is integrated using a Harten, Lax, Van Leer approximate Riemann solver for a diffusive limiter\footnote{Options \texttt{hll} and \texttt{min mod}, respectively, in the PLUTO setup.}. The initial magnetic fields are divergenceless, and this condition is maintained in the simulation with a hyperbolic/parabolic divergence cleaning technique \citep{ded}. A three-dimensional uniform spherical grid is used, with 128 radial cells, 48 cells in the polar angle $\theta\in [0,\pi]$ and 96 cells in the azimuthal angle $\phi\in [0,2\pi]$. Our computational domain consists of two spherical shells centered around the planet, which represent the inner $R_{\mathrm in}=1$ and outer boundary $R_{\mathrm out}=20$ of the system (and also of the computational box, expressed in the units of planetary radius). Parameters used in our simulations are given in the first two rows in Table~\ref{params}, where R$_{\mathrm NS}\sim 10$~km. The \texttt{SW} (\texttt{Speed, MagField, Dens, Temp}) for the pulsar environment are as given in \cite{Romanova05,Petri16}, with the density and the magnetic field in the pulsar wind decreasing linearly with the distance from the light cylinder. The realistic pulsar wind velocity in the simulations should be larger for about 0.3$\times 10^{10}$ cm/s, and the related magnetic field for about one order of magnitude, but at such velocities and magnetic fields our setup becomes unstable, and we relegate it to further work. In the setup for pulsar planets without intrinsic magnetic field there is no damping layer atop the planet, which was used in the Solar system setup, so there is no need for setting of the planet surface temperature \texttt{PlanTemp} or the radius of the damping shell R$_{\mathrm sw,cut}$. The same for the Alfv\'{e}n speed limit \texttt{AlfSpeedLimit}, which is in the special relativity setup by default limited to the speed of light. As elaborated in \cite{Mottez20}, if the pulsar wind is reaching the planet surface, the flow can be super-Alfv\'{e}nic and still produce the Alfv\'{e}n wings\footnote{Even without an atmosphere, a rocky planet under heavy bombardment of particles from the stellar wind can emanate enough material from the surface to produce visible aurora, as is the case on Mercury \citep{Varela16a}.}

 The magnetic Reynolds number in the simulation is due to the grid resolution, $R_{\mathrm m} = VL/ \eta \sim 10^3$, where $\eta$ is the magnetic diffusivity in the code. The characteristic length in the simulation is the planet radius $L\sim 10^8$~cm, and the characteristic velocity V$\sim 10^{10}$~cm~s$^{-1}$ is the pulsar wind velocity. Since we do not introduce physical resistivity, the reconnection of the magnetic field is driven by the numerical magnetic diffusivity, which was evaluated from numerical experiments with the same grid resolution in a simpler setup to be of the order of $\eta\sim 10^{12}$~cm$^2$~s$^{-1}$ \citep{Varela18}.
 
 The orientation of the IMF is defined by the direction of the unit vectors in the magnetic field. 
 In the cases without the intrinsic planetary magnetic field, which we study here, we set the conducting or ferromagnetic planetary surface with the inner boundary specified in a slightly different way than in the cases with magnetic planets described in \cite{Varela18}: the radial component of the planetary magnetic field is set to zero. The polar and azimuthal components are set as smoothly absorbed, without a gradient, by copying the values from the last active zone in the computational box, to the boundary ghost zone. In addition, in the ferromagnetic case, the azimuthal component of the magnetic field changes the sign at the inner boundary.

In the left panel in Fig.~\ref{schexo} is shown a schematic view of a star-planet system in a case with non-vanishing planetary magnetic field. The bow shock region is indicated by the increased density. The dynamic pressure from stellar wind bends the magnetic field lines on the day side of the planet and the field lines get elongated at the other side, forming a magnetotail. The IMF reconnects with the exoplanet magnetic field lines, which leads to the formation of the magnetopause. Aurora is a glow from the interaction of the planetary magnetosphere and stellar wind, or from the interaction of the inflowing particles from the stellar wind with the planetary surface moving through it. As in the solar wind, the plasma in the stellar wind carries the frozen-in magnetic field from the stellar surface, creating the IMF, which piles up near the planetary surface, inducing an increase in the local magnetic field there. 

The radio emission from Mercury was simulated in \cite{Varela16}, and results for the Earth-like aurora were presented in \cite{Varela18, Varela22}. Because of strong millisecond pulsar magnetic field, $B\sim10^{7-9}$~G, \citep[see e.g.,][]{1998pulsar} compared to solar-type stars\footnote{Millisecond pulsar magnetic field is large in comparison to the field of solar-type stars, but is orders of magnitude weaker than the field of newly born pulsars with $B\sim10^{12}$~G, \citep{BackerMilis82} or magnetars $B\sim10^{14}$~G, \citep{KaspiMagn17}.}, the pulsar planets could be stripped of their field, or it could be negligible. Even if there is some planetary magnetic field, because of the large dynamic and magnetic pressure in the pulsar wind the magnetopause would be at the surface of the planet, like in the cases of non-magnetic planets. Study of an aurora observed in the vicinity of non-magnetic planets, like Venus and Mars, is still at the beginning \citep{2022marsvenus}, but they could offer some insights for pulsar planets\footnote{The kind of glow which is seen from Venus magnetotail - which is produced by the solar magnetic field facing the non-magnetized obstacle \citep{2012venus} - is of particular interest: in a much stronger pulsar field, it could provide observable signature.}. The computation of radio emission could differ in detail from the one employed in the magnetized planets. Still, as the first step, we use the same method as previously for magnetized planets, only with the adjusted inner boundary conditions at the surface of the planet. In the next section, we provide the results of our pulsar-planet simulations and predict the observational signature of planets around millisecond pulsars.

\section{Results for non-magnetized planets around pulsars}\label{plplan}
\begin{table}
\caption{Parameters used in PLUTO setup file \texttt{pluto.ini} in our simulations for pulsar-planet setups with conductive and ferromagnetic planetary surfaces in comparison to Sun-Earth (CME) and Sun-Earth and Sun-Mercury (quiet) conditions. SW (Speed, MagField, Dens, and Temp) are setting the related initial values--in the Pulsar-planet case, SW corresponds to stellar or pulsar wind. PlanTemp sets the planetary temperature, and the Alfv\'{e}n speed is limited by the AlfSpeedLimit. The radii R$_{\mathrm in}$ and R$_{\mathrm sw,cut}$ set the inner boundary of the system and the radial position of the nose of the bow shock at the beginning of the simulation, respectively. The density floor is controlled by \texttt{dens\_{min}}=0.01$\times$ \texttt{SWDens}.}
\centering                          
\begin{tabular}{ c c c c c  c c c c c }        
\hline    
 Set-up & SWSpeed & SWMagField & SWDens & SWTemp & PlanTemp & AlfSpeedLimit & R$_{\mathrm in}$ &  R$_{\mathrm sw,cut}$\\
 & (cm~s$^{-1}$) & (G) & (g~cm$^{-3}$) & (K) & (K) & (cm~s$^{-1}$) & (R$_{\rm NS}$) &  (R$_{\rm NS}$) \\
\hline\hline
 Pulsar-planet (cond.) & $2.6\times 10^{10}$ & $2.5\times 10^{-3}$ & $5.0\times10^{-26}$ & $5.0\times 10^8$ & - & - & 1.0  & - \\
Pulsar-planet (ferro) & $2.6\times 10^{10}$ & $1.0\times 10^{-2} $ & $1.0\times 10^{-24}$ & $5.0\times 10^8$ & - & - & 1.0  & - \\ 
Sun-Earth (CME) & $1.0\times 10^8$ & $1.0\times 10^{-3}$ & $3.0\times 10^{-23}$ & $1.0\times 10^5$ & $1.0\times 10^3$ & $5.0\times 10^8$ & 3.0 & 6.0 \\
Sun-Earth (quiet) & $3.5\times 10^7$ & $5.0\times 10^{-5}$ & $6.0\times 10^{-24}$ & $4.0\times 10^4$ &$1.0\times 10^3$ & $5.0\times 10^8$ & 3.0 & 6.0  \\
Sun-Mercury (quiet) & $5.0\times 10^7$ & $1.5\times 10^{-4}$
& $2.0\times 10^{-23}$ & $8.0\times 10^4$ & $2.0\times 10^3$ & $1.0\times 10^8$ & 1.0 & 3.0 \\
\hline
\hline
\end{tabular}
\label{params}
\end{table}
On the planets orbiting around pulsars we can also expect auroras and simulations should provide predictions for their observation. In the right panel in Fig.~\ref{schexo} is presented the three-dimensional snapshot of the result in our pulsar-planet simulation, with the setup described in the previous section. We probed two different cases of planetary surface boundary conditions: conducting and ferromagnetic, with the parameters in the simulations as defined in Table~\ref{params}. For comparison, in the bottom rows are shown parameters used in already published simulations with the setup for the rocky planets in the Solar system. In our simulations for non-magnetized pulsar planets, the planetary temperature and the Alfv\'{e}n speed limit do not figure as parameters in the setup, so the former is assumed realistic for rocky planets, and the latter is implicitly equal to the speed of light.

The results in our simulations are shown in Fig.~\ref{simforsce3}. In the case of conducting exoplanetary surface, the pulsar magnetic field lines are connected with the planet's surface and the electric field lines are close to the surface of the planet. In the case of ferromagnetic exoplanetary surface, the electric field near the planet forms a dipole-like structure.

\subsection{Radio emission from planets around pulsars}\label{radioemis}
  \begin{figure}[ht!]
\plotone{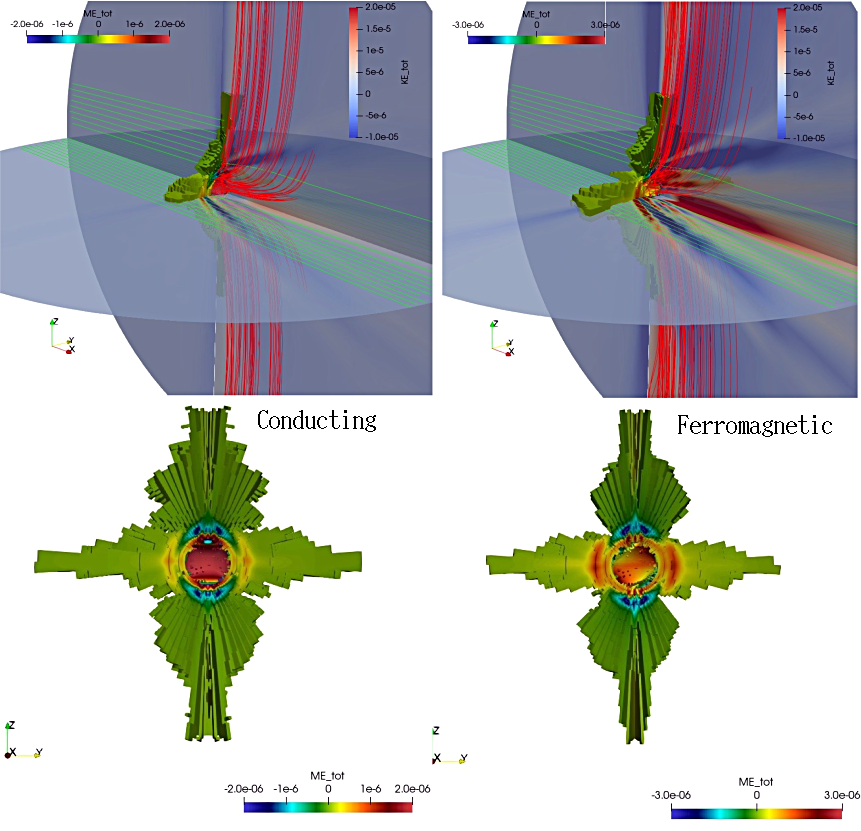}
      \caption{{\it Top panels}: the divergence of the Poynting flux {\bf ME}\_{\bf tot}$=\mathbf{E}\times\mathbf{B}/\mu_0$, where $\mu_0$ is the permeability of free space and the kinetic energy flux {\bf KE}\_{\bf tot}$=\rho\mathbf{v}|\mathbf{v}|^2/2$ term quantifying the pulsar wind dynamic pressure in the cases with the non-magnetic planet with a conducting and ferromagnetic surfaces, left and right panels respectively, are shown with the color graded isocontours. Color graded background shows the mass density $\rho$, red lines depict the magnetic field lines and green lines show the velocity streamlines of pulsar wind. {\it Bottom panels}: zoom into the radiative patterns in the same cases. Locations with the maximum radiated power are located in the middle of the planet dayside and in the base of the Alfv\'{e}n wings.}
         \label{nonmagrad}
   \end{figure} 
The electro-magnetic emission from the two simulated cases is shown in Fig.~\ref{nonmagrad}. The geometry of the radio emission is similar in both cases, only the intensity of the emission is slightly lower in the case with conducting planetary surface, probably because of the different details in the setup. The distribution of the local maxima in radio emission is different in each of the two configurations. In the case with conducting planetary surface, the maxima are located nearby the equatorial region and above the poles. With the ferromagnetic planetary surface, the maxima are located in-between the poles and the equator, and the local maxima are almost three times smaller than in the conducting case. The integrated radio emission in the conducting case is $3.65 \times 10^{12}$~W, while in the ferromagnetic case it is $1.14 \times 10^{13}$~W. From this we can predict the intensity of a signal from those planets measured on Earth. 

\subsection{Observation of aurora on pulsar planets}
Only a small fraction of pulsars, around 0.5 \%, are likely to host exoplanets of sufficient size to detect them (4 Earth masses or more). Even though they are rare, they could be detectable through the new generation of telescopes like LOFAR (Low-Frequency Array) \citep{LOFAR}, a large area radio telescope capable of detecting the radio signals emitted by the aurora of planets induced by their host stars. It is most sensitive at 150 MHz, where these interactions are expected to emit significant radiation \citep{Zarka07}. LOFAR was able to detect low-frequency radio waves that were predicted from a M-type dwarf GJ 1151 (or a planet around it) which is located 25 light-years from Earth \citep{2020Nat_Lofar_V}. This was, tentatively, the first signal detected from an extrasolar aurora. Together with already operating MeerKat facility, the upcoming new generation telescope square kilometer array (SKA), covering a wide range of low frequencies, could also serve as a tool for the detection of auroras on planets around pulsars.

The only case of star-planet interaction without non-thermal radio emission arises when both the planet and stellar wind are non-magnetized. In all the other cases, even without intrinsic planetary magnetic field, there can arise intense radio emission. Based on the observations of magnetized planets in the Solar system, the empirical generalized radiometric Bode's law (RBL) is employed to estimate the intensity of radio emission \citep{1984D,Zarka07}. The radiated power $P_{\mathrm rad}$ we compute as the partial transfer of the energy flow around the solid obstacle in the pulsar wind. Following \cite{Zarka07,Varela16}, the divergence of the magnetic Poynting flux is 
\begin{equation}
P_{\mathrm rad}=2\times 10^{-3}\int_V\mathbf{\nabla}\cdot{\mathbf{ME}}_{\mathbf{tot}}
dV=2\times 10^{-3}\int_V\mathbf{\nabla}\cdot\frac{(\mathbf{v}\times\mathbf{B})\times\mathbf{B}}{\mu_0}dV,
\label{pflux}
\end{equation}
where the factor $2\times 10^{-3}$ comes from the efficiency of dissipated power to radio emission conversion \citep{Zarka18}, and V is the volume enclosed between the center of the dayside of the planet and the magnetotail.

The density of the radio flux from a planet at a distance $d$ is then
\begin{equation}
\Phi=\frac{P_{\mathrm rad}}{\Omega d^2 \Delta\nu},
\label{phi1}
\end{equation}
with $\Omega=0.16$~sr being the solid angle of the beam of emitted radiation, which we set as ten times smaller than in the case of Jovian emission \citep{Zarka04}. It is a rough estimate; in the relativistic flow the emission region is expected to be more narrow than $\Omega=1.6$~sr estimate in the nonrelativistic Jovian case, as described in \cite{Mottez14,Mottez20}. The emission bandwidth $\Delta\nu$ is assumed to be the same as the maximum emission frequency \citep{Lynch18}.

The maximum emission frequency $\nu_{\mathrm max}$ and the characteristic plasma frequency $\nu_{\mathrm min}$ are computed as:
\begin{equation}
\nu_{\mathrm max}=\frac{e B_{\mathrm sw}}{2\pi m_{\mathrm e}}\sim 2.8~{\mathrm {MHz}}B_{\mathrm sw},\ \nu_{\mathrm min}=\sqrt{\frac{ne^2}{\pi m_{\mathrm e}}}\sim 8.98~{\mathrm {kHz}}\sqrt{n},
\label{nus1}
\end{equation}
with the electron charge $e$ and mass m$_{\mathrm e}$ and the pulsar wind magnetic field in the vicinity of the planet B$_{\mathrm sw}$ measured in Gauss and the plasma number density ${\mathrm n}$ in cm$^{-3}$ (\texttt{SWMagField} and \texttt{SWDens}/m$_{\mathrm e}$ in Table~\ref{params}, respectively). Below a lower limit of the frequency $\nu_{\mathrm min}$, emission is absorbed by the plasma-in our simulations, we are well above this limit near the planet's orbit around a pulsar. For the ground observation on Earth, the ionosphere, with a number density of $10^6$~cm$^{-3}$, absorbs all the frequencies below 10~MHz. In our simulations, which we could perform in the current setup only with an order or two smaller magnetic field than expected in the pulsar wind, we are below this limit.

In Table~\ref{observs} we list the obtained parameters from the simulations and give the predicted values for the radio emission flux for an observer on Earth in our two cases of pulsar planets with conducting and ferromagnetic planetary surfaces. 

For the integrated radio emission obtained in simulations, with a given distance of the planet and emission bandwidth, the values of density of radio flux for the pulsar-planet in the case with conductive planet, from the Eq.~\ref{phi1} the radio flux density $\Phi$ at a distance 750~pc is
\begin{equation}
    \Phi = \frac{3.65 \times 10^{12}}{0.16({\mathrm sr}) \cdot (750\cdot 3.1 \times 10^{16})^2\cdot 2.8\cdot 0.0025 (G) \times 10^6({\mathrm Hz})} \cdot 10^{26}~{\mathrm Jy} = 0.60~{\mathrm mJy} ,
\end{equation}
and similarly, for the pulsar-planet in the ferromagnetic case $\Phi=0.47$~mJy.
\begin{table}
\caption{Predicted intensity of the radio emission flux $\Phi$ for an observer on Earth in our two simulations with non-magnetized pulsar planet with conducting and ferromagnetic planetary surfaces. The planets are positioned at 750 pc, the distance to the planets around PSR 1257+12. We also show the predicted intensity of the emission at the typical distance of many known pulsars, 250 pc and at 100 pc.  In the last three columns we check if the predicted values are above the sensitivity and the frequency limit of the currently most sensitive instruments, LOFAR and MeerKAT, and the future SKA, for which minimal sensitivities are of the order of 0.1, 0.01 and 0.001 mJy, respectively. In brackets are given the results with expected larger value of B$_{\mathrm sw}$ equal to 7.4 G and 13 G in each of the cases, for which the $\Delta\nu$ are 20.1 and 36.4, respectively.}
\centering                          
\begin{tabular}{ c c c c c c c c c c}        
\hline    
 Set-up & $\Phi_{\mathrm a}(750)$ & $\Phi_{\mathrm b}(250)$ & $\Phi_{\mathrm c}(100)$ & P$_{\mathrm radio}$ & B$_{\mathrm sw}$ & $\Delta\nu$ & LOFAR & MeerKAT & SKA \\
 & (mJy) & (mJy) & (mJy) & (Wm$^{-2}$) & (G) & MHz & & & \\
\hline\hline
 Pulsar-planet (cond.) & 0.60 & 5.4 & 33.75 & $3.65\times 10^{12}$ & 0.0025 & 0.007 & NO(YES) & NO(YES) & NO(YES) \\
\hline 
Pulsar-planet (ferrom.) & 0.47 & 4.23 & 26.43 & $1.14\times 10^{13}$ & 0.01 & 0.028 & NO(YES) & NO(YES) & NO(YES) \\
\hline
\hline
\end{tabular}
\label{observs}
\end{table}
All the obtained values are beyond the current or planned instruments because of too low $\nu_{\mathrm max}$ for ground observations, but with the expected larger values of B$_{\mathrm sw}$ they would be of the same order, or larger, than the minimal sensitivity of the currently available or planned instruments. Our results suggest that auroras on planets around pulsars could be presently observable.

\section{Conclusions}\label{concl}
Our numerical simulations quantify the magnetospheric interaction between millisecond pulsars and planets around them. We investigate the potential for detecting planets around pulsars using radio emission. We focus on two scenarios with planets without intrinsic magnetic fields: pulsar planets with conducting and ferromagnetic planetary surfaces. The obtained density of radio flux for the Earth observer is of the order of 0.1~mJy to 30~mJy. The obtained frequency is above the minimal
frequency for which the emission would be absorbed by the plasma, but below the 10 MHz ionosphere cut-off. With the expected larger values of B$_{\mathrm sw}$, they are above this cut-off.

The results from our simulations show that planets around millisecond pulsars could be observable with present radio telescopes such as LOFAR, MeerKAT, and the future SKA, whose minimum sensitivities are of the order 0.1, 0.01 and 0.001 mJy, respectively. This exciting prospect opens up new possibilities for studying and understanding
the dynamics of pulsar systems and their planetary companions.

\section*{Acknowledgements}
The work in CAMK was funded by a Polish NCN grant no. 2019/33/B/ST9/01564. M\v{C} acknowledges the Czech Science Foundation (GA\v{C}R) grant No.~21-06825X and the support by the International Space Science Institute (ISSI) in Bern, which hosted the International Team project \#495 (Feeding the spinning top) with its inspiring discussions. JV acknowledges project 2019-T1/AMB-13648 funded by the Comunidad de Madrid. Authors thank to Profs. W. Klu\'{z}niak, P. Haensel and J. Szul{\'a}gyi for informative discussions and comments, and to the anonymous referee for constructive suggestions.

\section*{Data Availability}
The simulation data associated with this article will be made available by the corresponding author upon reasonable request.

\bibliography{pulsplan}{}
\bibliographystyle{aasjournal}

\end{document}